\documentclass{PoS}
\usepackage{latexsym}
\usepackage{bm}
\usepackage{amsfonts}
\usepackage{amssymb}
\usepackage{amsmath}
\usepackage{array}
\usepackage{capt-of}

\title{Cosmic-ray transport in the heliosphere with {\tt HelioProp}}

\ShortTitle{CR transport in the heliosphere with {\tt HelioProp}}

\author{\speaker{Andrea Vittino}\\
        Physik-Department T30d, Technische Universit{\"a}t M{\"u}nchen,
James Franck-Str. 1, D-85748, Garching, Germany\\
        E-mail: \email{andrea.vittino@tum.de}}
                
\author{Carmelo Evoli\\
        Gran Sasso Science Institute, Viale Francesco Crispi 7, 67100  L'Aquila, Italy\\
        E-mail: \email{carmelo.evoli@gssi.it}}
        
\author{Daniele Gaggero\\
        GRAPPA, University of Amsterdam, Science Park 904, 1098 XH Amsterdam, Netherlands\\
        E-mail: \email{d.gaggero@uva.nl}}        

\abstract{Before being detected at Earth, charged cosmic rays propagate across the Solar System and undergo interactions with the turbulent solar wind and with the heliospheric magnetic field. As a result, they are subject to a series of processes that include diffusion, convection, energy losses and drifts, which significantly affect the shape and the intensity of the cosmic-ray fluxes at low energies. Here we illustrate how all these mechanisms can be realistically modelled with {\tt HelioProp}, our public tool designed to treat cosmic-ray transport through the heliosphere in a charge-dependent way. We present a detailed description of the features of the code and we illustrate in a quantitative way the effects that the propagation in the heliosphere can have on the different cosmic ray species with a particular emphasis on the antiparticle channels relevant for dark matter indirect detection.}

\FullConference{35th International Cosmic Ray Conference --- ICRC2017\\
		10--20 July, 2017\\
		Bexco, Busan, Korea}

\begin{document}

\section{Introduction}
\label{sec:introduction}
Before reaching the site of their detection at Earth, Cosmic Rays (CRs) propagate across the Solar System, where they are subject to the interaction with the turbulent solar wind and with the Heliospheric Magnetic Field (HMF). The ensemble of effects that arise as the result of these interactions can significantly alter CR intensity, in particular at low energies. This mechanism goes typically under the name of solar modulation (we address the reader to \cite{2013LRSP...10....3P} for a review). 

An accurate description of solar modulation represents a necessary ingredient in the study of CR properties, also in connection with the search for exotic CR sources, such as Dark Matter (DM). This is the reason why we have developed {\tt HelioProp}, a tool to model the transport of CRs in the heliosphere by means of the realistic three-dimensional model based on the Stochastic Differential Equation (SDE) technique described in \cite{2011ApJ...735...83S,2012Ap&SS.339..223S}. 

Results based on an earlier version of the code have been reported in \cite{Maccione:2012cu,Fornengo:2013osa,Fornengo:2013xda,Gaggero:2013nfa}. The analyses described here have been conducted with a novel version of {\tt HelioProp} which is currently under development and which will be released soon as a completely open-source package. In addition, {\tt HelioProp} will be part of the {\tt Dragon2} project \cite{Evoli:2016xgn}. Together, the two codes aim at describing CR transport from the source to the observer under conditions that are as general and realistic as possible. 

This proceeding is organised as follows: in Section~\ref{sec:solarmod} we briefly describe the most important ingredients that characterise the physics of solar modulation, while in Section~\ref{sec:numerics} we illustrate the main features of the numerical method that is adopted in {\tt HelioProp} to model CR transport in the heliosphere. In Section~\ref{sec:tests} we present some results obtained with the code, in particular for what concerns the modelling of CR proton and electron fluxes and the impact of solar modulation on DM searches. Lastly, in Section~\ref{sec:conclusions} we report our conclusions.
 
\section{The physics of solar modulation}
\label{sec:solarmod}
Following the model proposed by Parker \cite{1958ApJ...128..664P}, the HMF is typically assumed to possess a spiral structure. If we consider a spherical coordinate system $\{r,\theta,\phi\}$ with origin at the Sun location, the HMF can be written as:
\begin{equation}
\vec{B}(r,\theta,\phi) = A_c B_0 \left( \frac{r_0}{r}\right)^2 \left[ \hat{r} - \left( \frac{\Omega r \,\mathrm{sin}\theta}{V_{\mathrm{sw}}}\right) \hat{\phi} \right],
\label{eq:HMF}
\end{equation} 
where $r_0$ = 1 AU and $\Omega$ is the differential rotation rate of the Sun, while $V_{\mathrm{sw}}$ represents the velocity of the solar wind, which is a radial flow continuously directed outward from the Sun. All the results that are shown in this work have been derived by assuming for the radial and latitudinal dependence of $V_{\mathrm{sw}}$ the expression given in \cite{2014SoPh..289..391P}.  Concerning the parameters that define the normalization of the HMF, $B_0$ can be related to the intensity of the HMF at Earth $B_e$,
while $A_c$ is the HMF polarity, defined as $A_c = \pm H (\theta - \theta')$ where the sign is known to shift with a period of roughly 11 years, while $H$ is the Heaviside step function and $\theta'$ is the angular extent of the wavy Heliospheric Current Sheet (HCS). This quantity is defined as follows:
\begin{equation}
\theta' = \frac{\pi}{2} + \mathrm{sin}^{-1} \left[ \mathrm{sin} \alpha \, \mathrm{sin} \left( \phi - \phi_0 + \frac {\Omega r}{V_{\mathrm{sw}}}\right)\right]
\label{eq:tilt_angle}
\end{equation} 
where $\phi_0$ is an (arbitrary) constant azimuthal phase and $\alpha \in [0,\pi/2]$ is the HCS tilt angle \cite{1992sws..coll..191H}, which is a quantity that varies according to solar activity and reaches its lowest values ($\alpha \approx 10^\circ$) in solar minima periods.
 
The transport of Cosmic-Rays (CRs) across the heliosphere is described in terms of a transport equation \cite{1965P&SS...13....9P}:
\begin{equation}
\frac{\partial f}{\partial t} = - \vec{V}_{\mathrm{sw}} \cdot \nabla f  + \langle \vec{v}_{\mathrm d}\rangle \cdot \nabla f + \nabla \cdot (\vec{K_s} \cdot \nabla f) + \frac{1}{3} (\nabla \cdot \vec{V}_{\mathrm{sw}} ) \frac {\partial f}{\partial {\mathrm{ln}} R} 
\label{eq:TPE}
\end{equation}
where $R=pc/eZ$ is the rigidity of the CR particle under consideration while $f$ is its distribution function, related to the CR differential intensity $j$ by $f = R^2j$. The four terms on the r.h.s. are used to model, respectively, convection under the influence of the solar wind, drifts, spatial diffusion and adiabatic energy changes. 

Concerning the process of CR drifts, it is modelled through the drift velocity (averaged with respect to the pitch angle) $\langle \vec{v}_{\mathrm d}\rangle$ that appears in Eq.~(\ref{eq:TPE}). Since drifts are caused both by the interaction of CRs with the gradient and curvature of the HMF and by the change in the field polarity at the crossing of the HCS, the drift velocity can be written as the sum of two components: $\langle \vec{v}_{\mathrm d}\rangle = \langle \vec{v}_{\mathrm d}\rangle_{\mathrm{GC}} + \langle \vec{v}_{\mathrm d}\rangle_{\mathrm{HCS}}$. These two components are implemented in {\tt HelioProp} according to the prescriptions given in \cite{2011ApJ...735...83S,2012Ap&SS.339..223S}, to which we address the reader for additional details. It is important to remark that, following \cite{2011ApJ...735...83S}, we include in our model a progressive weakening of drifts towards small rigidities. This is done by multiplying the drift velocity by the reduction factor:

\begin{equation}
f(R)_{\mathrm{drifts}} = \frac{\left(\frac{R}{R_d}\right)^2}{1+\left(\frac{R}{R_d}\right)^2}
\label{eq:drifts}
\end{equation}

Concerning the process of CR spatial diffusion, it is related to the tensor $\vec{K_s}$, which represents the symmetric part of the diffusion tensor. More precisely, one can work in the reference frame of the average HMF (frame that is denoted here with the superscript $B$) and decompose the total diffusion tensor into a symmetric and an antisymmetric component as follows:   
\begin{equation}
K^{B}_{ij} = K^{B,S}_{ij} + K^{B,A}_{ij} =
\begin{pmatrix}
k_{\parallel} & 0 & 0 \\
0 & k_{\perp,r} & 0 \\ 
0 & 0 & k_{\perp,\theta}  \\ 
\end{pmatrix}
+
\begin{pmatrix}
0 & 0 & 0 \\
0 & 0 & k_A \\ 
0 & -k_A & 0  \\  
\end{pmatrix},
\label{eq:k}
\end{equation}
where $k_{\parallel}$ is the diffusion coefficient along a direction parallel to the one of the HMF, while $k_{\perp,r}$ and $k_{\perp,\theta}$ are the diffusion coefficients along the radial and polar perpendicular directions. The antisymmetric part of the diffusion tensor $K^{B,A}_{ij}$ accounts for drifts and is already included in Eq.~(\ref{eq:TPE}) in terms of the drift velocity $\vec{v}_{\rm d} = \nabla \times \left(K_A \frac{\vec{B}}{|B|} \right)$. On the other hand, as already said, the symmetric tensor $K^{B,S}_{ij}$, once translated into its counterpart in spherical coordinates $K^{S}_{ij}$, enters the transport equation to model spatial diffusion. As one can from Eq.~(\ref{eq:k}), diffusion is totally described once that the coefficients $k_{\parallel}$, $k_{\perp,r}$ and $k_{\perp,\theta}$ are determined. As typical for the most realistic descriptions of solar modulation, {\tt HelioProp} features a fully anisotropic diffusion, which means that it works under the assumption that $k_{\parallel} \ne k_{\perp,r} \ne k_{\perp,\theta}$. We express the three diffusion coefficients in terms of the Mean Free Paths (MFPs) in the associated direction $k_{i} = v/3 \lambda_i$ with $v$ being the CR velocity. We follow a common approach (see \cite{2013LRSP...10....3P} for example) and define the parallel mean-free-path as a broken power-law with break at rigidity $R_k$ and indices $a$ and $b$ below and above the break: 
\begin{equation}
\lambda_{\parallel} = \lambda_0 \left(\frac{1\, \mathrm{nT}}{B}\right) \left( \frac{R}{1\, \mathrm{GV}} \right)^{a} \left(\frac{\left( \frac{R_k}{1\, \mathrm{GV}} \right)^c + \left( \frac{R_k}{1\, \mathrm{GV}} \right)^c }{1 + \left( \frac{R_k}{1\, \mathrm{GV}} \right)^c}\right)^{\frac{b-a}{c}}.
\label{eq:kpar_break}
\end{equation}
As for the MFPs along the perpendicular directions, {\tt HelioProp} allow the user to implement any generic function. In this work, we assume: 
\begin{equation}
\lambda_{\perp,r} =f_\perp \lambda_{\parallel} , \quad\quad\lambda_{\perp,\theta} =f_\perp \lambda_{\parallel} H(\theta) , 
\label{eq:kperp}
\end{equation}
with the function $H(\theta)$ being defined as in \cite{2013LRSP...10....3P}. 
\section{Numerical solution of the transport equation}
\label{sec:numerics}

{\tt HelioProp} models solar modulation by numerically solving Eq.~(\ref{eq:TPE}). This is done by following the SDE technique described in \cite{2011ApJ...735...83S,2012Ap&SS.339..223S}, which was originally proposed in the context of CR solar modulation in \cite{1999AdSpR..23..505Y,1999ApJ...513..409Z}. In brief, this method is realised by writing the transport equation as a backward Kolmogorov equation: 
\begin{equation}
-\frac{\partial j}{\partial s} = \sum_i \left( A_i\frac{\partial j}{\partial x_i}\right) + \frac{1}{2} \sum_{i,k} \left( C_{ik} \frac{\partial^2 j}{\partial x_i \partial x_k }\right)
\end{equation} 
where $j$ is the CR differential intensity, while the parameter $s$ is a backward time, related to the standard time $t$ by $t = t_{\mathrm{fin}} - s$, where $ t_{\mathrm{fin}}$ is the final time. The It\={o}'s lemma \cite{ito1944} states that the above equation is equivalent to a set of stochastic processes: 
\begin{equation}
dx_i = A_i(x_i)ds + \sum_j B_{ij} (x_i) dW_i
\label{eq:SDE}
\end{equation}
where we have used $C = B^\mathrm{T} B$, while $x_i = \{r,\theta,\phi,E\}$ and $W_i$ is a Wiener process, i.e.~a stochastic variable for which the increments $\Delta W_i = w_i - w_{i-1}$ are independent one from the other and each of them has a gaussian probability distribution with zero mean and standard deviation equal to $\Delta s_i = t_i - s_{s-1} $. We address the reader to \cite{2011ApJ...735...83S,2012Ap&SS.339..223S} for the explicit form of the stochastic processes described by Eq.~(\ref{eq:SDE}).

Under a practical point of view, when following this approach one models solar modulation by combining the backward trajectories in the phase space of $N$ pseudo-particles. More precisely, these pseudo-particles are injected at Earth at $s=0$ with phase space coordinates $x_{\mathrm{Earth}}$ and then they are back-propagated through the heliosphere by following the stochastic processes described by Eq.~(\ref{eq:SDE}) until they reach the boundary of the heliopause (HP). 
The CR flux at Earth is then obtained by averaging the Local Interstellar (LIS) spectra evaluated at the coordinates $x_{\mathrm{HP}}$ that the pseudo-particles have at the HP: 
\begin{equation}
j_{\mathrm{Earth}} (x_{\mathrm{Earth}}) = \frac{1}{N} \sum_{k=1}^N j_{\mathrm{LIS}}(x_\mathrm{HP})
\end{equation}  


\section{Results}
\label{sec:tests}
\subsection{Proton and electron fluxes}
\label{sec:protons}


The first test that we perform consists in comparing the results of {\tt HelioProp} with the proton and electron fluxes measured by PAMELA over short time intervals (one month for the protons, six months in the case of electrons) along the period of minimal solar activity and negative polarity of the HMF that goes from 2006 to 2009 \cite{2013ApJ...765...91A,2015ApJ...810..142A}\footnote{Because of space limitations, here we restrict ourselves to 4 electron datasets out of the 7 released by the PAMELA Collaboration. }. 

For each data-taking period that we consider, we fix the tilt angle $\alpha$ and the HMF at Earth $B_e$ to the values reported in \cite{2014SoPh..289..391P,2015ApJ...810..141P}. We are thus left with a set of free parameters that include the quantities $\lambda_0$, $a$, $b$, $c$ and $R_k$ that appear in Eq.~(\ref{eq:kpar_break}), the factor $f_\perp$ of Eq.~(\ref{eq:kperp}) and the rigidity $R_d$ of Eq.~(\ref{eq:drifts}). For simplicity, we impose that electrons and protons share the same values for both $R_d$ and $f_\perp$. In particular, we find that good fits to data can be found by assuming $R_d$ = 0.32 GV, which is the same value used in \cite{2013ApJ...765...91A}, and $f_\perp$ = 0.021, that is very close to the commonly used value 0.02 (see \cite{2013LRSP...10....3P}).  We assume the parallel MFP $\lambda_\parallel$ of protons to have a linear dependence on the CR rigidity as in the model of \cite{2012Ap&SS.339..223S}. This means that, for protons, $a$ = $b$ = 1 (and therefore $c$ and $R_k$ do not play any role, as clear from Eq.~(\ref{eq:kpar_break})). In the case of electrons, we find that we are able to reproduce the data behaviour by assuming $a$ = 0, $b$ = 1.55, $c=3.5$. A rigidity-dependence of the electron MFP in terms of a  broken power-law of this kind (i.e. with a flattening at small rigidities) appears to be in a qualitative agreement with previous analyses of electron data, as the one reported in \cite{2015ApJ...810..141P}. With this considered, the only parameters that are left free to vary from one dataset to the other are $\lambda_0$ (for both protons and electrons) and $R_k$ (only for electrons). Concerning the proton and electron LIS fluxes, we adopt the parameterizations based on Voyager 1 and PAMELA data presented in \cite{refId0}.  

\begin{figure}[t]
\center
\scriptsize
\begin{tabular}{|p{2cm}|p{2cm}|p{2cm}|p{2cm}|p{2cm}|}
\hline
\multicolumn{5}{|c|}{{\bf protons} ($a$ = $b$ = 1, $f_{\perp}$ = 0.021) }\\
\hline
& {\bf 2006} & {\bf 2007} & {\bf 2008} & {\bf 2009} \\
\hline
$\lambda_{\parallel}$ [AU] & 0.015  & 0.018 & 0.020 & 0.023 \\
\hline
\hline
\multicolumn{5}{|c|}{{\bf electrons} ($a$ = 0, $b$ = 1.55, $c=3.5$, $f_{\perp}$ = 0.021, $R_{d}$ = 0.32 GV)}\\
\hline
& {\bf 2006} & {\bf 2007} & {\bf 2008} & {\bf 2009} \\
\hline
$\lambda_{\parallel}$ [AU] & 0.019 & 0.021 & 0.026 & 0.029 \\
\hline
$R_k$ [GV] & 0.44 & 0.43 & 0.48 & 0.50 \\
\hline

\end{tabular}
\captionof{table}{Parameters of the solar modulation models used to reproduce PAMELA electron and positron data, as detailed in Section~\ref{sec:protons} of the text. The values of $\lambda_{\parallel}$ are the ones at 100 MV at Earth.}
\label{tab:protons_electrons}
\centering
\includegraphics[width=0.41\textwidth]{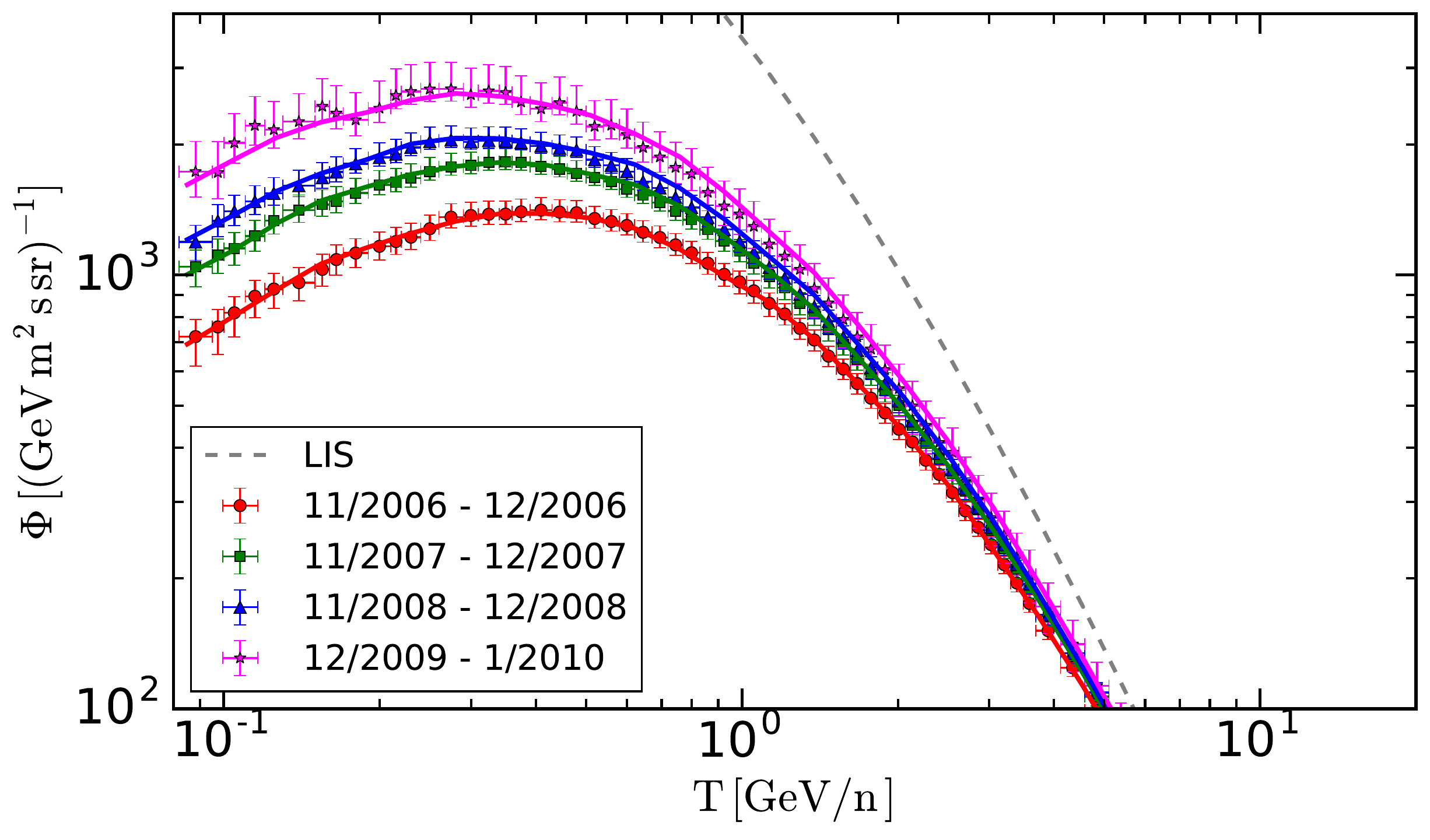}
\hspace{0.5cm}
\includegraphics[width=0.425\textwidth]{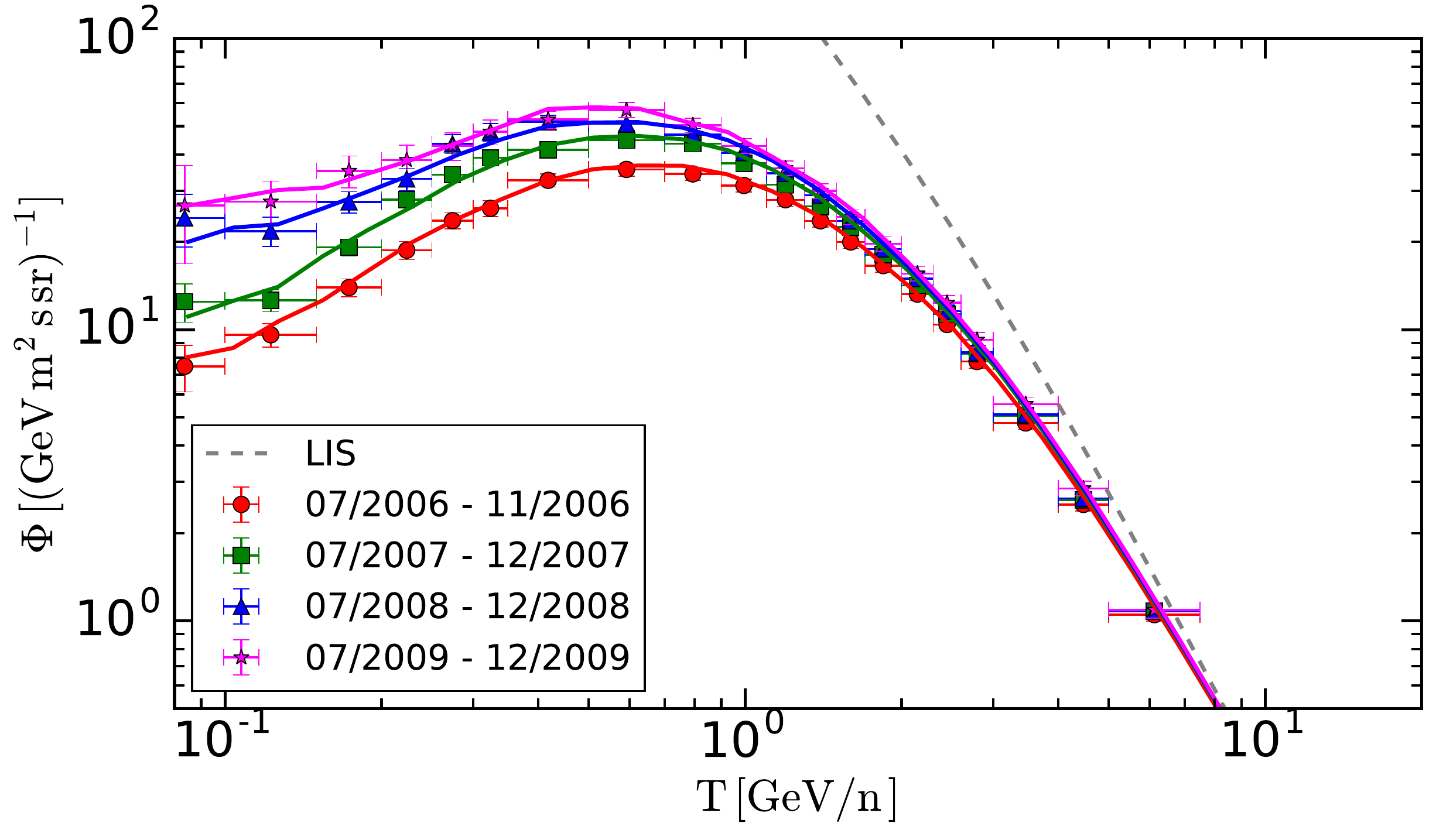}\\
\includegraphics[width=0.41\textwidth]{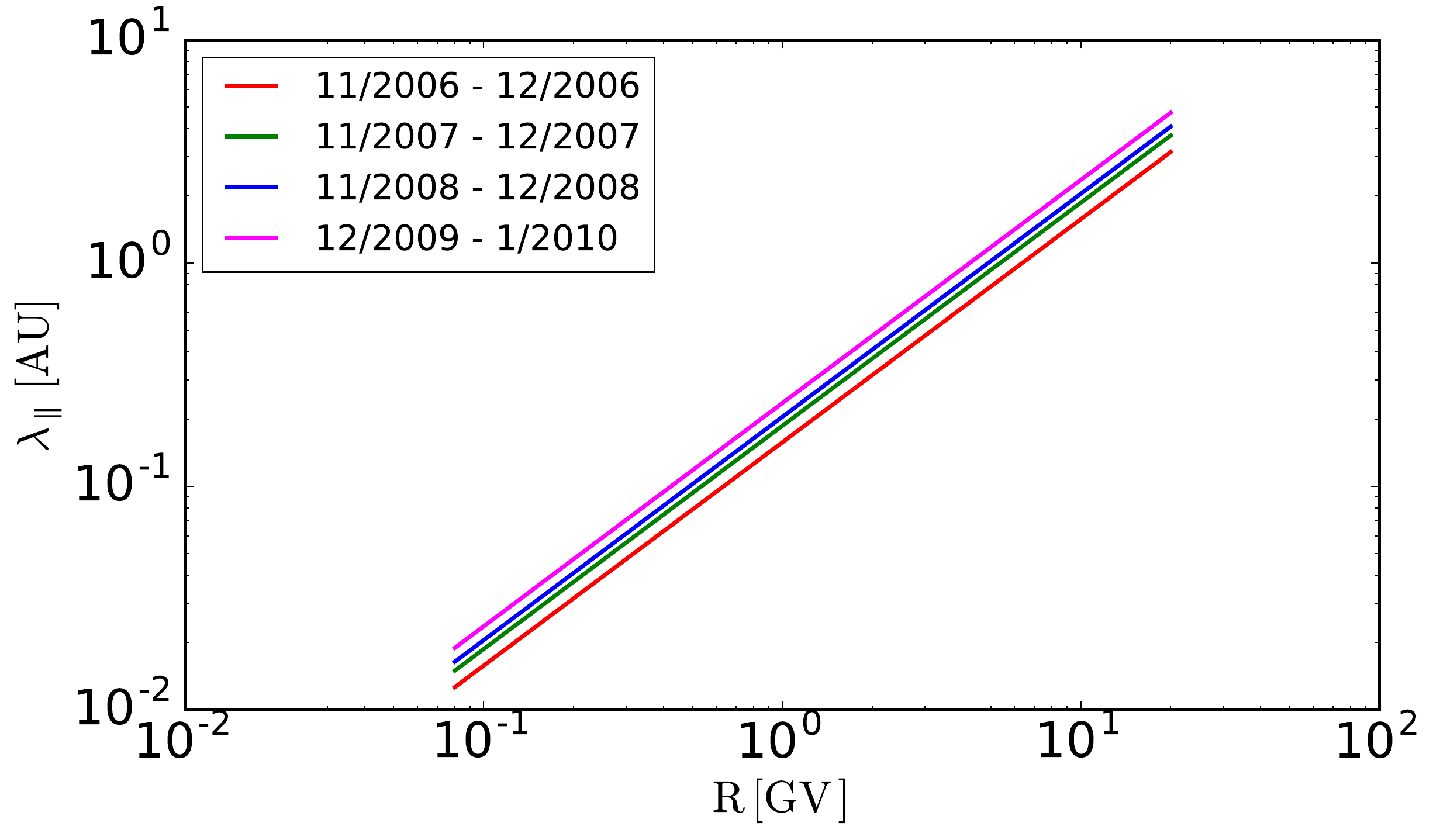}
\hspace{0.5cm}
\includegraphics[width=0.41\textwidth]{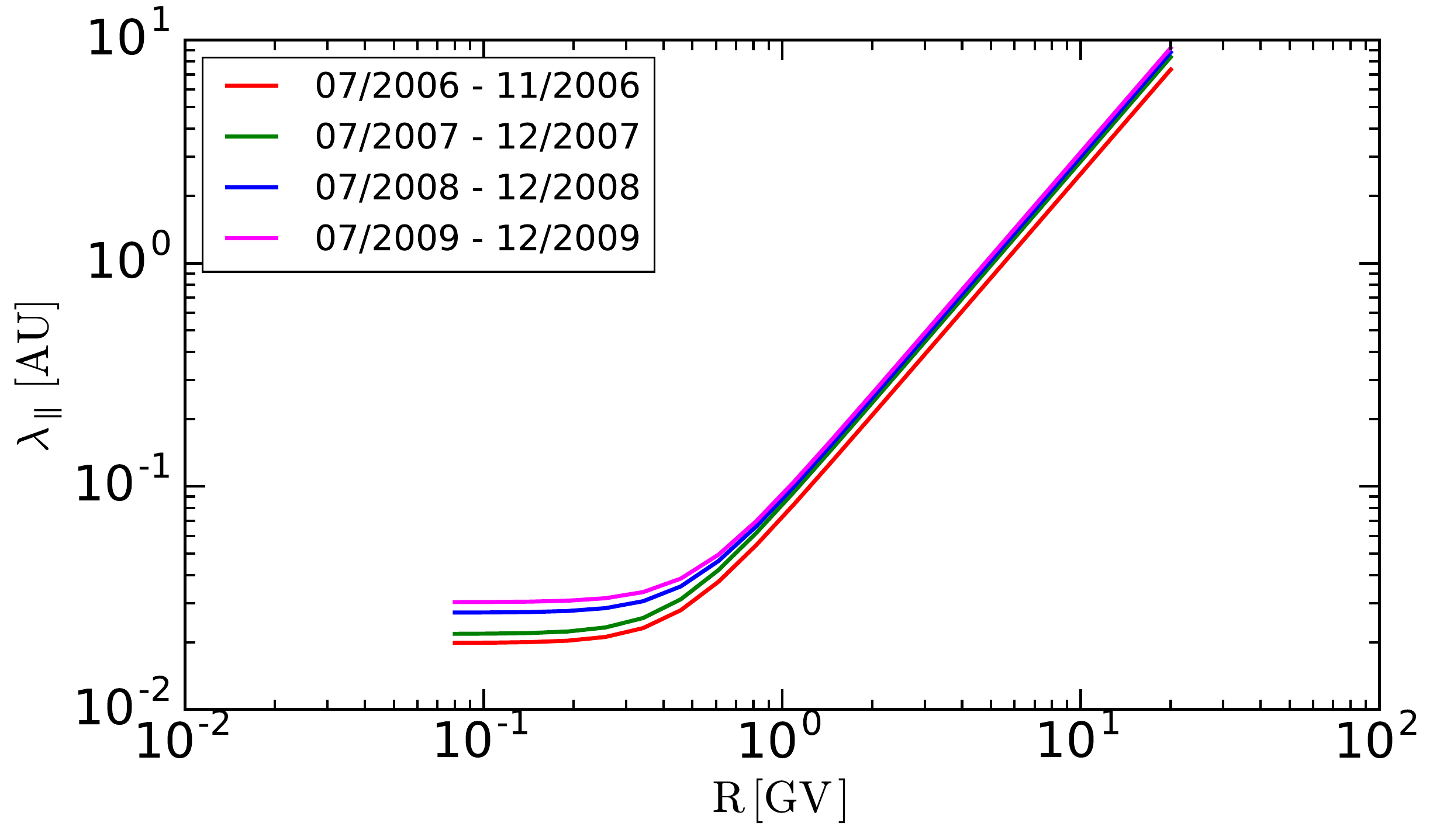}
\captionof{figure}{In the top row, the data points represent PAMELA datasets (as reported in the boxed insets), while the lines correspond to the proton (left panel) and electron (right panel) fluxes as predicted by {\tt HelioProp}, according to the solar modulation model detailed in the text and with parameters as in Table~\ref{tab:protons_electrons}. In the bottom row the parallel MFPs for protons (left panel) and electrons (right panel) are shown.}
\label{fig:res_CR}
\end{figure}

A summary of the relevant parameters used in this analysis is reported in Table~\ref{tab:protons_electrons}. Results are reported in Fig.~\ref{fig:res_CR}, In the top row we show PAMELA data compared with the best-fit configurations obtained within the framework of our model, while in the bottom row the corresponding parallel MFPs at Earth are shown as a function of the CR rigidity. 

As it can be seen, the model that we are using can reproduce the observed behaviour of both protons and electrons. It is important to remark that the approach followed here is data-driven, since we fix the values of the the parameters characterising CR diffusion and drifts by fitting PAMELA data; as discussed in Section~{\ref{sec:solarmod}}, {\tt HelioProp} allows for a large freedom in the definition of the solar modulation setup and therefore one could instead follow a more theoretically motivated strategy and adopt for these parameters the results obtained in the context of turbulence transport models (see, e.g.,  \cite{0004-637X-642-1-230, 2013ApJ...772...46E}). 

\subsection{Antiprotons and consequences for DM searches}
\label{sec:DM}

In the context of indirect DM searches in charged CRs, it is customary to treat solar modulation in terms of the force field approximation \cite{1968ApJ...154.1011G}. Here we provide a simple example of what can be the impact of adopting a more realistic solar modulation setup, as the one implemented in {\tt HelioProp}. We do this by comparing predictions of DM-generated antiproton fluxes obtained with {\tt HelioProp} and with the force-field approach. Analyses along these lines have been already performed with the previous version of {\tt HelioProp} and can be found in \cite{Fornengo:2013osa} (for the case of antiprotons) and in \cite{Fornengo:2013xda} (for the case of antideuterons). 

Since we are dealing with an exotic source of antiprotons contributing mostly at low energies, we cannot tune the solar modulation parameters by directly using antiproton observations and we must rely on proton data alone. Obviously, we have to consider a proton and an antiproton datasets derived from observations performed over the same time interval. This is the case of the PAMELA datasets presented in  \cite{2010PhRvL.105l1101A,2011Sci...332...69A}. We find that we can fit the PAMELA proton dataset by using the same setup described in Section~\ref{sec:protons} (with $\lambda_{\parallel}$ = 0.016 AU at 100 MV at Earth). In the framework of the force field approximation, a comparably good fit is provided by assuming a force-field potential of 450 MV. 

\begin{figure}[t]
\centering
\includegraphics[width=0.40\textwidth]{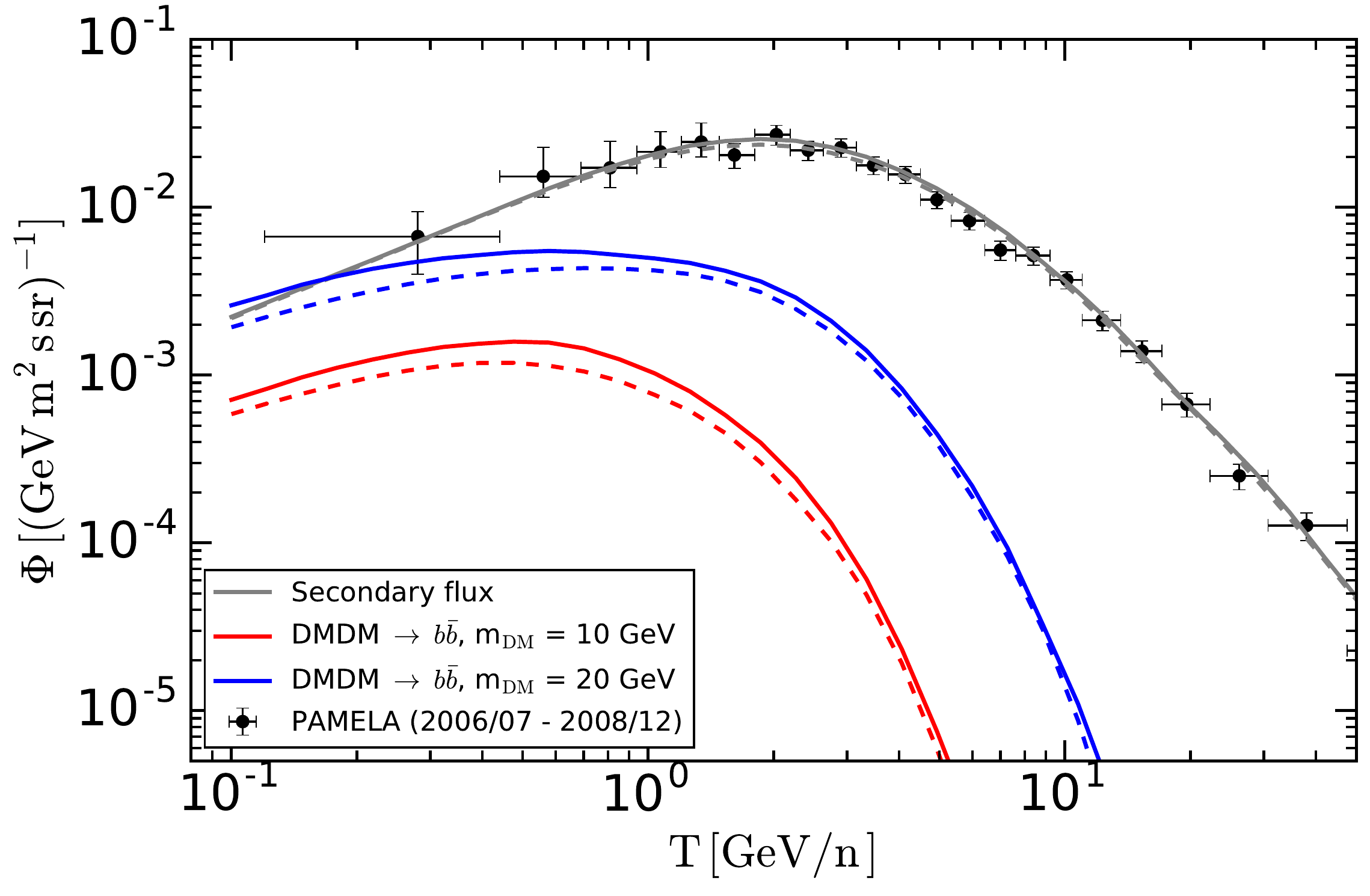}
\hspace{0.7cm}
\includegraphics[width=0.40\textwidth]{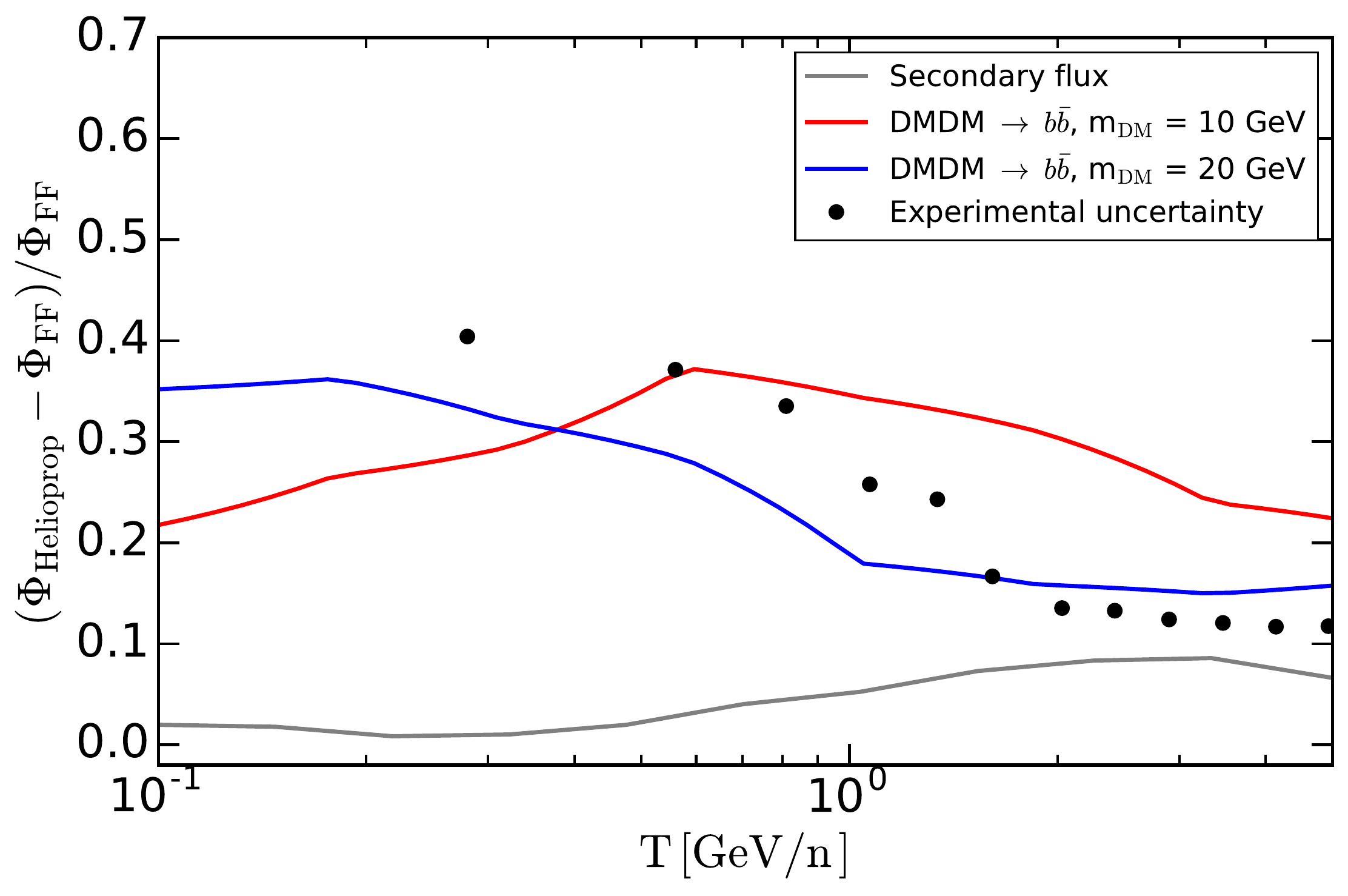}
\caption{In the left panel, the antiproton fluxes predicted by {\tt Helioprop} (solid lines) and by the force field approximation (dashed lines) are shown, together with PAMELA data points. The annihilation cross sections used for the computation of the DM fluxes are $\langle \sigma v \rangle = 5\times10^{-27}$ cm$^3$s$^{-1}$ for m$_{\mathrm{DM}}$ = 10 GeV and $1.5\times10^{-26}$ cm$^3$s$^{-1}$ for m$_{\mathrm{DM}}$ = 20 GeV. In the right panel, the fractional difference between the two modulation models is shown for the different fluxes under consideration, together with the experimental accuracy. This quantity is defined at every PAMELA data point as as the ratio between the experimental error (on the flux, computed as the average between the upper and lower error) and the measured flux.}
\label{fig:res_DM}
\end{figure}

As for antiprotons, we adopt a LIS derived from a {\tt DRAGON} run performed within the {\sc KRA} propagation model (see \cite{2010APh....34..274D} for details about this propagation setup)\footnote{It is important to remark that the {\tt DRAGON} run has been performed by tuning the proton injection spectrum in order to match the LIS used for protons in this work.}. The antiproton fluxes at Earth obtained by modulating this LIS with {\tt HelioProp} and with the FF approach are shown in Fig.~\ref{fig:res_DM}: as one can see, the difference between the two modulation setups is minimal (below 10\%) in the case of the antiproton flux produced by spallation processes. On the contrary, if one considers the antiproton flux produced by the annihilation of light DM, as we do here by assuming a DM particle with mass 10 or 20 GeV that annihilates into $b\bar{b}$, the difference is larger and can reach the 35\%. This is shown in the right panel of Fig.~\ref{fig:res_DM}, where it is also shown that such a distance can be larger than the experimental uncertainty. This proves that when deriving constraints to DM properties from low-energy CR data, a realistic treatment of solar modulation is necessary.    

\section{Conclusions}
\label{sec:conclusions}
We have presented here the new version of {\tt HelioProp} the numerical code designed to model solar modulation in a three-dimensional and charge-dependent way by using the SDE approach described in \cite{2011ApJ...735...83S,2012Ap&SS.339..223S}.  {\tt HelioProp} is designed to be part of the {\tt Dragon2} project \cite{Evoli:2016xgn}, but will also be released soon as an independent and fully open source package. 

We have briefly illustrated the main features of the code, under a physical and numerical point of view. In addition, we have presented a couple of basic tests aimed at investigating its performances and potentialities. In particular, we have shown that {\tt HelioProp} is able to reproduce the observed behaviour of low-energy CR electrons and protons and we have illustrated how it can be an important tool for a more accurate treatment of DM indirect detection.   

\bibliography{helioprop}


\end{document}